\documentclass[submitting]{nst}

\usepackage{subfigure,dcolumn}
\usepackage[T2A,T1]{fontenc}
\usepackage[russian,english]{babel}

\usepackage{listings}
\usepackage{tikz,xcolor,hyperref}

\definecolor{lime}{HTML}{A6CE39}
\DeclareRobustCommand{\orcidicon}{
	\begin{tikzpicture}
	\draw[lime, fill=lime] (0,0) 
	circle [radius=0.16] 
	node[white] {{\fontfamily{qag}\selectfont \tiny ID}};
	\draw[white, fill=white] (-0.0625,0.095) 
	circle [radius=0.007];
	\end{tikzpicture}
	\hspace{-2mm}
}
\foreach \x in {A, ..., Z}{%
	\expandafter\xdef\csname orcid\x\endcsname{\noexpand\href{https://orcid.org/\csname orcidauthor\x\endcsname}{\noexpand\orcidicon}}
}

\usepackage{lineno}
\linenumbers

\begin{document}

\title{Underground Laboratory JUNA Shedding Light on Stellar Nucleosynthesis}
\author{Toshitaka Kajino\orcidA{}}\thanks{Email: kajino@buaa.edu.cn}
\affiliation{School of Physics, and International Research Center for Big-Bang Cosmology and Element Genesis, Beihang University,
37 Xueyuan Rd., Haidian-district, Beijing 100083 China}
\affiliation{Division of Science, National Astronomical Observatory of Japan,
2-21-1 Osawa, Mitaka, Tokyo 181-8588, Japan}
\affiliation{Graduate School of Science, The University of Tokyo, 
7-3-1 Hongo, Bunkyo-ku, Tokyo 113-033, Japan}


\maketitle

{\bf Abstract: 
Extremely low background experiments to measure the key nuclear reaction cross sections of astrophysical interest are carried out at the world's deepest underground laboratory, Jingping Underground laboratory for Nuclear Astrophysics (JUNA). High precision measurements provide reliable information to understand the nucleosynthetic processes in celestial objects and help solve mysteries on the origin of atomic nuclei which were discovered in the first generations of Pop. III stars in the universe and the meteoritic SiC grains in the solar system.} 


Nuclear astrophysics is a growing interdisciplinary research field of nuclear physics and astrophysics. We seek for the origin and evolution of hundreds of atomic nuclei in the universe to understand the evolution of stars and galaxies. 
In addition to the events of the centuries, SN1987A and GW170817, a huge number of complementary data have been accumulated progressively from the high-resolution spectroscopic observations of old faint stars and the precise meteoritic analysis of isotopic anomalies and correlations among them. 

\begin{figure*}
\includegraphics
[width=1.0\linewidth]
{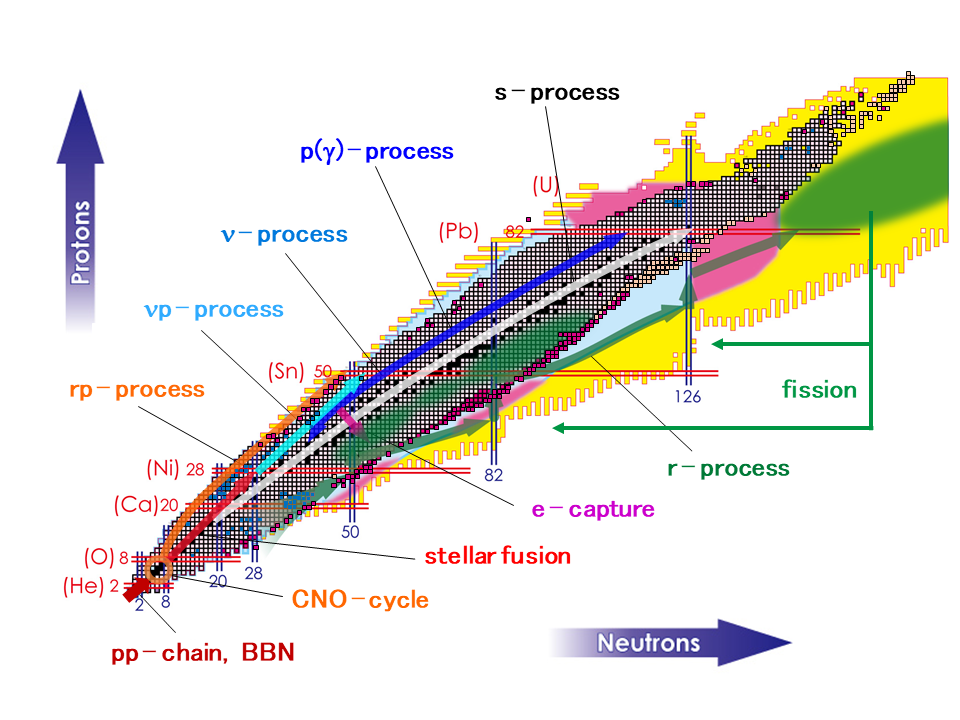}
\caption{\label{fig:nucl_chart} Nuclear chart and the known nucleosynthetic processes in the universe. Filled and open black boxes are the stable and radioactive isotopes, respectively, whose lifetimes were measured. Red boxes are the newly synthesized isotopes and blue boxes are those with known masses. Isotopes in the yellow region are predicted theoretically. See ref.~\cite{APPS2021} for more details.}
\end{figure*}

The origin of atomic nuclei is classified into three typical astrophysical sites and epochs in the cosmic evolution~\cite{Bertulani2016}; Big-Bang nucleosynthesis in the first 3-10 min of the early universe for the production of light mass nuclei, thermonuclear and explosive nucleosyntheses in stars for the production of intermediate-to-heavy mass nuclei after the galaxies formed, and Galactic cosmic-ray interactions with interstellar medium in space for the production of rare abundant nuclei. It is a long standing challenge in nuclear astrophysics~\cite{B2FH1957} to identify the key nuclear reactions and precisely determine their reaction rates~\cite{APPS2021} in several nucleosynthetic processes shown in Figure~\ref{fig:nucl_chart}. 

Several nucleosynthetic paths among them such as rapid-neutron capture process (r-process) involve many radioactive unstable nuclei, which are difficult to study experimentally because of their short lifetimes and low beam intensities. 
Even should the science target be the measurement of the cross sections of stable nuclei, most of the cosmic or stellar nucleosynthesis occurs at temperatures $T = 10^8-10^9$K under the extreme conditions. Although these temperatures are among the higher temperatures of various astronomical phenomena, the relevant energy scale is as low as $E = 10-100$ keV.
This energy scale is too low to directly measure the reaction cross sections of charged particles in the laboratories because extremely small Coulomb penetrability significantly reduces the cross section\cite{Rolfs}.  

Another foremost difficulty is the huge cosmic-ray background that prevents the precise measurement of small reaction cross sections. It is particularly critical to reduce the cosmic-ray background to detect neutral particles like photons and neutrons in order to measure the reliable reaction rates of radiative-capture $(\alpha,\gamma)$, (p,~$\gamma$) and neutron emission ($\alpha$,~n) reactions,
which take the keys to all nucleosynthetic processes in Fig.~\ref{fig:nucl_chart}.

\begin{figure*}
\includegraphics
[width=1.0\linewidth]
{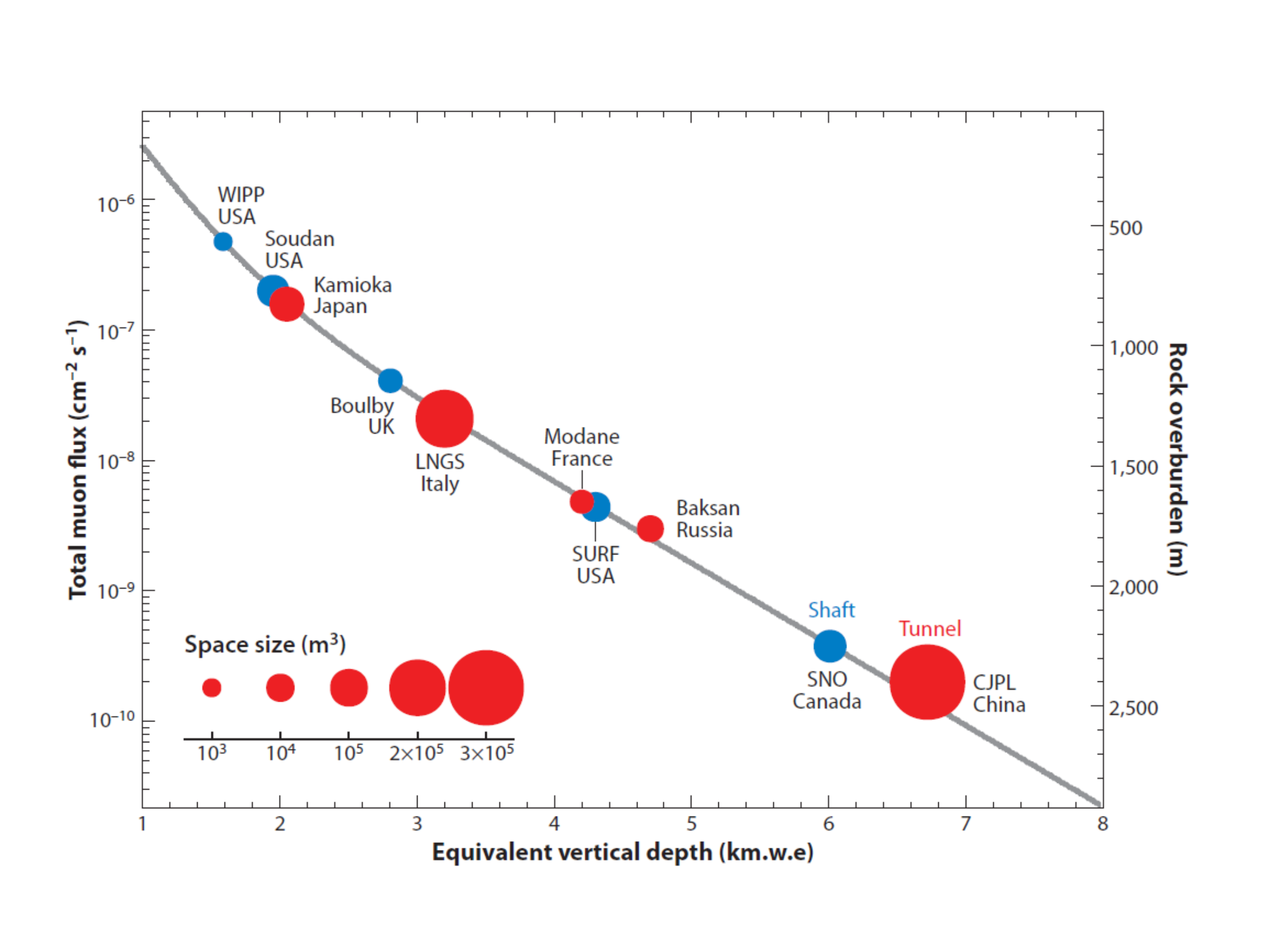}
\caption{\label{fig:JUNA_depth} The measured residual muon fluxes in key underground facilities over the world, which are consistent with predicted values (grey line). The sizes of the circles correspond to laboratory space by volume; red or blue denotes access by road tunnels or shafts, respectively.}
\end{figure*}

JUNA Collaboration is directed by CIAE, jointly supported by NSFC, CNNC and CAS, and has established a quite unique underground facility equipped with high-intensity accelerator for the direct measurements of extremely small cross sections at astrophysical low energies. JUNA locates 2400 meters deep at the China JinPing underground Laboratory complex (CJPL-II) established in 2014 that is the deepest underground laboratory around the world (Fig.~\ref{fig:JUNA_depth}). The biggest advantage is the rock with thickness of several kilometers to shield the cosmic rays and corresponding backgrounds. Since the first beam was delivered in Dec. 2020, experiments on several key reactions have been carried out successively with the high intensity accelerator at the ultra-low background environment.

A day-one campaign at JUNA was dedicated to the experiment on 
$^{19}$F(p,~$\alpha\gamma)^{16}$O at very low energies $E_{cm}$ = 72.4 - 344 keV covering the Gamow window~\cite{Zhang2021,Zhang2022PRC}, for the first time. Fluorine is one of the most mysterious monoisotopic elements whose predicted abundance is subject to large uncertainty of the reaction rates. It is claimed to have multiple origins, i.e. the asymptotic giant branch stars (AGB stars) in relatively recent epoch with solar metallicity~\cite{Cristallo2009}, and the core-collapse type II supernova (cc-SN II) in the early galaxy at low metallicity~\cite{Woosley1988}. Theoretical uncertainties are reduced significantly by the JUNA experiment, and the breakout condition of carbon–nitrogen–oxygen cycle (CNO cycle) in AGB stars is more precisely constrained. The SN $\nu$-process nucleosynthesis of $^{19}$F is also revisited by taking account of the flavor oscillation effects to infer nucleosynthetic constraint on still unknown mass hierarchy~\cite{Ko2022}. 

Another day-one campaign also was successfully done for  $^{25}$Mg(p,$\gamma$)$^{26}$Al which plays an important role in the production of $^{26}$Al ($\tau_{1/2}=7.17\times10^5$ y) in massive srars~\cite{Su2022}. The 1.809 MeV $\gamma$-ray emitted from $^{26}$Al was observed by the Gamma-ray Satellite INTEGRAL that provides distribution and total accumulated mass of $^{26}$Al in the Milky Way. The estimated frequency of cc-SN II and Ib/c turned out to be 1.9 $\pm$ 1.1 events per century~\cite{Diehl2006}, which gives the firm basis to the theoretical studies of Galactic chemical evolution of r-process elements~\cite{Yamazaki2022}. JUNA Collaboration team determined the reaction rate for $^{25}$Mg(p,$\gamma$)$^{26}$Al at high accuracy by measuring the resonance parameters at 92 keV and contributes to establish this scheme of estimating the SN event rate.
 
The production mechanism of $^{40}$Ca observed in the oldest ultra-metal-poor star~\cite{Keller2014} is an unsolved mystery in the recent astronomical observations. In the standard stellar evolution model, most of $^{19}$F produced by the hot CNO cycle would be recycled back to $^{16}$O by the $^{19}$F(p,~$\alpha)^{16}$O reaction, and therefore the flow could not reach the production of $^{40}$Ca. This scenario, however, is subject to error bars of its competing reaction rate for $^{19}$F(p,~$\gamma)^{20}$Ne that breaks the hot CNO cycle as the temperature increases. JUNA Collaboration team directly measured this reaction down to $E_{cm}$ = 186 keV, and found that the key resonance located at
225 keV contributes to a large enhancement in thermonuclear reaction rate for $^{19}$F(p,~$\gamma)^{20}$Ne~\cite{Zhang2022Nature}. Theoretical calculation of hydrostatic burning in
Pop. III stars using the new rate results in the $^{40}$Ca abundance consistent with those observed in the oldest known ultra-metal-poor stars~\cite{Zhang2022Nature}.

JUNA Collaboration team challenged 
$^{13}$C($\alpha$,~n)$^{16}$O reaction and the direct measurement of the cross section in the range of $E_{cm}$ = 240 keV$-$1.9 MeV was successfully done, removing large uncertainty in the previous experimental data~\cite{Gao2022}. Since this reaction and $^{22}$Ne($\alpha$,~n)$^{25}$Mg are known to be the major neutron source for the s-process in AGB stars, precise determination has long been wanted to predict reliable nuclear abundances from iron to $^{209}$Bi theoretically. JUNA experiment covers Gamow window for the intermediate-neutron capture process (i-process) that has recently been identified as a new process in metal-deficient AGB stars and even the collapsar, a major r-process site of massive star collapsing to a black hole~\cite{He2023}. Comprehensive theoretical studies of the s-, i- and r-processes in multiple astrophysical sites are underway with JUNA's new precise data. 

Quite recently, JUNA Collaboration team has reported a new experimental result~\cite{Wang2023} on $^{18}$O$(\alpha,~\gamma)^{22}$Ne. They succeed in the precise determination, for the first time, of the energy $E_{lab} = 474.0 \pm 1.1$ keV, spin-parity $J^{\pi} = 1^-$ and $\gamma$-width $\omega_{\gamma} = 0.25 \pm 0.03$ $\mu$eV of the most effective resonance dominating the total reaction rate at Gamow window. $^{22}$Ne is produced via $^{14}$N$(\alpha,~\gamma)^{18}$F$(e^- \nu)^{18}$O$(\alpha,~\gamma)^{22}$Ne that takes the key to identify the conversion efficiency from CNO cycle, whose main product is $^{14}$N, to the advanced burning stage where the s-process occurs by the use of neutrons produced via subsequent reaction $^{22}$Ne($\alpha$,~n)$^{25}$Mg in both AGB and supergiant stars. 
The improved JUNA data also contribute on probing the unknown origin of isotopic abundance ratio $^{21}$Ne/$^{22}$Ne found in meteoritic stardust SiC grains. The so-called mainstream components of SiC grains arise from AGB stardust. Therefore, both $^{18}$O$(\alpha,~\gamma)^{22}$Ne and its competing reaction $^{18}$O$(\alpha$,~n)$^{21}$Ne affect strongly the resultant isotopic ratio. The present improved JUNA data on the former reaction remove the major uncertaity in this isotopic ratio. The observed $^{21}$Ne/$^{22}$Ne ratios in meteoritic stardust SiC grains are used to constrain the physical conditions of nuclear burning processes which depend on the initial mass of parent AGB star and other astronomical parameters in modeling stellar evolution. 

JUNA Phase-I has almost finished its successful performance of high precision measurements of several key reactions at the lowest cosmic-ray background. JUNA Phase II is underway and the following reactions would be the next targets: $^{12}$C($\alpha$,~$\gamma$)$^{16}$ to solve the still poorly known stellar evolution and explosion mechanism of massive stars and the mystery of massive black-hole mass-gap~\cite{Farmer2019}; $^{12}$C($^{12}$C,~$\alpha$)$^{20}$Ne and $^{12}$C($^{12}$C,~p)$^{23}$Na to identify still unknown heat source of X-ray Superburst, and to clarify double-degenerate explosion mechanism of SN Ia~\cite{Mori2019}; $^3$He($\alpha,\gamma$)$^7$Be and $^3$H($\alpha,\gamma$)$^7$Li to solve overproduction problem of the Big-Bang lithium, and to precisely constrain the neutrino-mixing parameters from missing solar neutrino flux~\cite{Kajino2023}; $^{11}$B($\alpha$,~n)$^{14}$N, $^{15}$N($\alpha,~\gamma$)$^{19}$F, $^{17}$O(n,~$\alpha$)$^{14}$C and its reverse reaction to elucidate the physical conditions of $\alpha$-rich freezeout and the r-process nucleosynthesis in magneto-hydrodynamic jets from cc-SN II, collapsars and binary neutron star mergers~\cite{Kajino2019,Yamazaki2022}. JUNA is expected to continuously produce relevant nuclear data to enrich our knowledge of stellar and cosmic nucleosyntheses to solve many mysteries in astronomy and astrophysics. 


\end{document}